# Micromagnetic Modeling of the Magnetization Behavior of NiMnGa FMSMA's.

A.A. Likhachev and Yu.N. Koval

*Institute for Metal Physics, National Academy of Sciences, 36 Vernadsky Str., 03680,Kiev, Ukraine,* e-mail: alexl@imp.kiev.ua

Abstract

The model calculations of the magnetization curves for the internally twinned NiMnGa ferromagnetic shape memory alloy are performed at the different volume fractions of twin variants. The method is based on the direct minimization of our new micromagnetic free energy model of FMSMA's taking into account both the magnetic anisotropy energy and the magnetostatic energy contributions associated with the laminated twin microstructure. The effect of the magnetostatic energy is discussed in comparison with some early models, where the magnetostatic energy was completely ignored.

Introduction

Ni-Mn-Ga based Ferromagnetic Shape Memory Alloys (FMSMA's) have a unical ability to show an extremely large magnetic field induced deformation effects which is about of 30-50 times larger compared to the best known ordinary magnetostrictive materials [1]. First, these effects were discovered in two different nonstochiometric ferromagnetic martensitic phases of NiMnGa alloy ( 6% in 5M [2] and then 10% in 7M) [3]. It has been found that the strain mechanism in FMSMA's is based on the twin boundary motion and the resulting redistribution between two twin related variants A and B of the martensitic phase, which easy magnetization axes are perpendicular each to other [8-12, 17-21]. Both the multiple twin microstructure and the magneto-optical image of $180^0$ magnetic domains inside of the twin bands of the 5M martensite of NiMnGa alloy are shown in Fig.1 below.

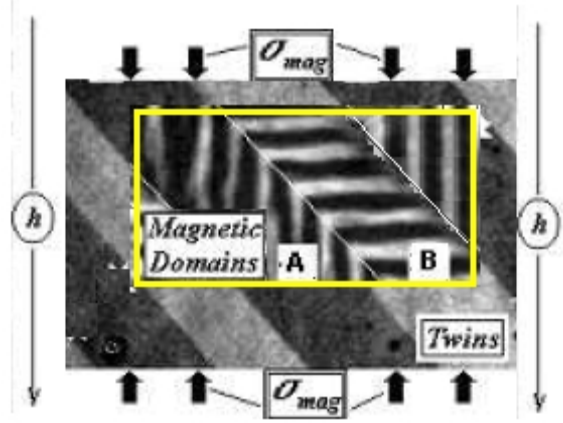

**Fig.1**. The twin microstructure consisting of two martensite variants A and B and the magneto-optical image ( at the insert ) of $180^0$ magnetic domains within the internally twinned 5M martensite of NiMnGa alloy.

It has been proved that the twinning process in NiMnGa is driven by a macroscopic magnetostrictive force developed during the magnetization of this material. According to [8], this force can be generally defined as follows:

$$\sigma_{mag}(h, x_A) = -\frac{\partial}{\partial \varepsilon} g_{mag}(h, x_A); \quad \text{where}, \quad \varepsilon = \varepsilon_0 x_A \qquad (1)$$

Here, $\sigma_{mag}(h, x_A)$ is the magnetostrictive force component acting along the external magnetic field on the transversal unit cross-section ( as shown in **Fig.1** ). Here, $g_{mag}(h, x_A)$ is the magnetic free energy per unit volume dependent on the magnetic field $h$ and the volume fraction $x_A$ occupied by the martensite variant A. The large strain $\varepsilon = \varepsilon_0 x_A$ which is developed in FMSMA's increases proportionally to the volume fraction $x_A$ and achieves its maximal value $\varepsilon_0$, which is a crystallographic constant dependent on the martensitic crystal lattice parameters. In NiMnGa alloys it can take 6% in 5M and 10% in 7M martensitic phases, correspondingly [8-12].

Our present results obtained on the basis of micromagnetic theory proposed in [13-16] show that along with the magnetic anisotropy and Zeeman's energies, the magnetostatic energy also plays an important role and produces a special coupling effect between the twin related martensite variants. It is caused by the demagnetizing effects both on the surface of the MSM material and also on the internal twin boundaries. As a result, the magnetostatic energy produces the nonlinear dependence of the magnetic free energy and the magnetic driving force both on the magnetic field and on the volume fractions of twin variants.

**Magnetic free energy model**

As follows from our micromagnetic free energy model proposed in [13-16] for the Ferromagnetic Shape Memory Materials, it consists of tree terms. The magnetic anisotropy energy is the first one:

$$F_{Ani+Zee} = K_U \left( x_A (\mathbf{m}_{A\perp}/M_s)^2 + x_B (\mathbf{m}_{B\perp}/M_s)^2 \right) - \mathbf{Mh} \quad (2)$$

Here and later, $x_A$ and $x_B$ are the volume fractions of the twin related martensite variants A and B, respectively; $K_U$ and $M_s$ are the uniaxial magnetic anisotropy constant and saturation magnetization of the MSM material, correspondingly. Typically, $K_U \approx 1.7 \bullet 10^5 \, J/m^3$ and $4\pi M_s = 0.65T$ in NiMnGa-based FMSMA's. Everywhere below, $\mathbf{m}_A$ and $\mathbf{m}_B$ denote the local magnetizations averaged over the fine magnetic domain microstructure within both twin variants A and B. The subscript sign "$\perp$" means a magnetization component perpendicular to the local easy magnetization direction of the corresponding twin variant.

The second term in Eq.2 is a so-called Zeeman's energy describing the effect of the external magnetic field. It is dependent on the total magnetization $\mathbf{M} = x_A \mathbf{m}_A + x_B \mathbf{m}_B$ averaged over the twin microstructure and the external magnetic field $\mathbf{h}$.

The third term represents the magnetostatic energy per unit volume:

$$U_{mag} = -\frac{1}{2}(x_A \mathbf{m}_A \mathbf{h}_A + x_B \mathbf{m}_B \mathbf{h}_B) \quad (3)$$

It also depends on the local demagnetizing field values $(\mathbf{h}_A)$ and $(\mathbf{h}_B)$ within the twin bands A and B, respectively. Similar to the macroscopic magnetization value $\mathbf{M} = x_A \mathbf{m}_A + x_B \mathbf{m}_B$ we can introduce the macroscopic demagnetizing field $\mathbf{H}^D = x_A \mathbf{h}_A + x_B \mathbf{h}_B$. Then, using identities $x_A = 1 - x_B$ and $x_B = 1 - x_A$ we can obtain:

$$\mathbf{h}_A = \mathbf{H}^D + x_B (\mathbf{h}_A - \mathbf{h}_B) \quad \text{and} \quad \mathbf{h}_B = \mathbf{H}^D - x_A (\mathbf{h}_A - \mathbf{h}_B) \quad (4)$$

According to a well known from the magnetism theory boundary conditions, both the normal component of the magnetic induction and the tangential components of the magnetic field must be continuous at the twin boundary interfaces. So, $(\mathbf{h}_A + 4\pi \mathbf{m}_A)_\mathbf{n} = (\mathbf{h}_B + 4\pi \mathbf{m}_B)_\mathbf{n}$ and $(\mathbf{h}_A)_\mathbf{t} = (\mathbf{h}_B)_\mathbf{t}$, respectively. Finally, these boundary conditions give us an important linear relationship between the local demagnetizing field and magnetization jumps at the twin boundaries:

$$\mathbf{h}_A - \mathbf{h}_B = -4\pi \mathbf{n} (\mathbf{m}_A - \mathbf{m}_B)_\mathbf{n} \quad (5)$$

Here $\mathbf{n}$ is a unit normal vector at the twin boundaries oriented at $45^0$ to the field direction: $n_x = n_y = 1/\sqrt{2}$; $n_z = 0$. There is also a well known linear relationship between the macroscopic demagnetizing field and the average magnetization of any ferromagnetic material:

$$\mathbf{H}^D = -4\pi \hat{\mathbf{D}} \mathbf{M} \quad (6)$$

Here, $\hat{\mathbf{D}}$ is a so-called demagnetizing matrix dependent only on the shape of a particular ferromagnetic sample. This matrix is always positively defined an has the unit its spur value. In a particular case, if the sample's shape is symmetric with respect to all reflections and inversion of **x**, **y**, and **z** coordinates in **Fig.1**, then a corresponding demagnetizing matrix will be diagonal one with all positive matrix elements satisfying the unit spur value: $Sp\,\hat{\mathbf{D}} = D_{xx} + D_{yy} + D_{zz} = 1$.

Finally, the magnetostatic energy can be represented as follows:

$$U_{Mag} = 4\pi\left(\frac{1}{2}\mathbf{M}\hat{\mathbf{D}}\mathbf{M} + \frac{1}{2}x_A x_B (\mathbf{m}_A - \mathbf{m}_B)_\mathbf{n}^2\right) \qquad (7)$$

$$\mathbf{M} = x_A \mathbf{m}_A + x_B \mathbf{m}_B \qquad (8)$$

Therefore, the magnetostatic energy consists of two terms. The first one is caused by an interaction between the magnetic charges induced at the external sample interface. In the similar way, the second one represents the interaction between the magnetic charges induced at the twin boundaries.

Finally the total magnetic free energy is:

$$F_{Mag}(\mathbf{m}_A, \mathbf{m}_B) = F_{Ani+Zee}(\mathbf{m}_A, \mathbf{m}_B) + U_{Mag}(\mathbf{m}_A, \mathbf{m}_B) \qquad (9)$$

It should be minimized with respect to the local magnetization variables $\mathbf{m}_A$ and $\mathbf{m}_B$, satisfying some additional restrictions:

$$(\mathbf{m}_A)^2 \leq M_s^2;\quad and \quad (\mathbf{m}_B)^2 \leq M_s^2 \qquad (10)$$

It is convenient to introduce four dimensionless magnetization vector components: $\mathbf{v} = (v_{Ax}, v_{Ay}, v_{Bx}, v_{By})$ instead of $\mathbf{m}_A$ and $\mathbf{m}_B$ as follows:

$$\mathbf{m}_A = (m_{Ax}, m_{Ay}) = M_s(v_{Ax}, v_{Ay});\quad \mathbf{m}_B = (m_{Bx}, m_{By}) = M_s(v_{Bx}, v_{By}) \qquad (11)$$

After that the magnetic free energy can also be represented in a dimensionless form:

$$F_{Mag}(\mathbf{m}_A, \mathbf{m}_B) = 4\pi M_s^2 f_{Mag}(\mathbf{v}) \qquad (12)$$

Here,

$$f_{Mag}(\mathbf{v}) = k_u \left( x_A v_{Ax}^2 + x_B v_{By}^2 \right) + \frac{1}{2} \left[ D_{xx} \left( x_A v_{Ax} + x_B v_{Bx} \right)^2 + D_{yy} \left( x_A v_{Ay} + x_B v_{By} \right)^2 \right] +$$
$$+ \frac{1}{4} \left[ x_A x_B \left( (v_{Ax} - v_{Bx}) + (v_{Ay} - v_{By}) \right)^2 \right] - h_0 \left( x_A v_{Ay} + x_B v_{By} \right) \quad , \quad (13)$$

Here, $h_0$ is a dimensionless parameter characterizing the external magnetic field: $h = 4\pi M_s h_0$ with- $4\pi M_s = 0.65T$. One more material parameter- $k_u = K_U / 4\pi M_s^2$ represents the dimensionless magnetic anisotropy energy constant. This dimensionless magnetization free energy must be minimized within the four-dimension region:

$$v_{Ax}^2 + v_{Ay}^2 \leq 1; \qquad v_{Bx}^2 + v_{By}^2 \leq 1 \quad (14)$$

**Minimizing procedure and results**

In this section we will consider the minimization free energy problem in one particular case when the FMSMA sample has a thin cylindrical shape, which is magnetized parallel to its long axis. It's well known, that if the cylinder is much longer than its diameter then the demagnetizing factor along its y-axis becomes zero, so as two other components along x- and z-axes will be equal ½. We will also choose the material parameters for the saturation magnetization and the anisotropy constant typical for the 5M-martensitic phase of NiMnGa FMSMA samples. So, we should take $4\pi M_s = 0.65T$ and $2K_U / M_s = 0.66T$ and define: $k_u = K_u / 4\pi M_s^2 = 0.51$.

Unfortunately, in presence of the magnetostatic interaction between the twin variants a minimization procedure should be done in the complex four dimension space area, where $\mathbf{m}_A^2(h, x_A) \leq M_s^2$ and $\mathbf{m}_B^2(h, x_A) \leq M_s^2$. So, it can be done only by using some numerical methods.

Practically, one can use the Nelder–Mead method (Nelder, J.A. and Mead, R. (1965)) which is a commonly applied numerical method used to find the minimum or maximum of a function in a multidimensional space. It is a direct search method which can be applied to nonlinear optimization problems for which derivatives may not be known.

Generally, it is expected, that a magnetization process in FMSMA's consists from three stages. At the first one, the absolute values of both local magnetizations will increase, remaining less of their saturation magnetization values: $|\mathbf{m}_A| \leq M_s, |\mathbf{m}_B| \leq M_s$. The magnetic field $h$ will increase from its zero value until the variant A becomes first fully saturated $|\mathbf{m}_A| = M_s$ at some critical field value: $h = h_A^S$.

At the second stage the magnetization within the variant A will remain constant $|\mathbf{m}_A| = M_s$ and may change its value by rotation only. At the same time, the magnetization within the variant B will continue to increase remaining less of its saturation value: $|\mathbf{m}_B| \leq M_s$, as the magnetic field $h$ increases from $h = h_A^S$ until the variant B becomes also fully saturated $|\mathbf{m}_B| = M_s$ at some second critical field value: $h = h_B^S$.

At the final third stage, both absolute local magnetization values will be remaining constant: $|\mathbf{m}_A| = M_s, |\mathbf{m}_B| = M_s$, as the magnetic field $h$ increases from $h = h_B^S$. During this stage both the magnetizations will change their values by rotation until their directions will become completely parallel to the external magnetic field direction.

In order to obtain some analytic results caused by the magnetostatic energy we can use a simplified micromagnetic energy model following from the general approach [13, 14], assuming the local magnetizations $\mathbf{m}_A(h, x_A)$, $\mathbf{m}_B(h, x_A)$ to be exactly parallel to the external field and neglecting their perpendicular components. Therefore, the magnetic free energy will consist of two parts: a zero order contribution- $g_{mag}^0(h, x_A)$, obtained from Eq.2, and also of the first order term- $g_{int}(h, x_A)$ following from the Eq.7.

$$g^0(h, x_A) = x_A(-hm_A) + x_B\left(K_u\left(\frac{m_B}{M_s}\right)^2 - hm_B\right) \quad (15)$$

$$g_{int}(h, x_A) = \frac{1}{2} 4\pi D (x_A m_A + x_B m_B)^2 + \frac{1}{2} x_A x_B 4\pi (S)(m_A - m_B)^2 \quad (16)$$

Here, $S = (ne_h)^2 = 0.5$ and $n, e_h$ are unit vectors parallel to the twin boundary normal and the magnetic field direction, respectively.

As follows from Eq.15, a zero order contribution $g_{mag}^0(h, x_A)$ is linearly dependent on the volume fraction of both twin variants. Its minimization is well known and widely discussed in different publications [8-10]. In this zero order approach the local magnetizations of both variants are parallel to the external field and grow linearly $m_A(h) = M_s(h/h^A)$, $m_B(h) = M_s(h/h^B)$ till their full saturation at $h = h^A$ and $h = h^B$. Here, the local saturation fields are denoted as $h^A = 4\pi D M_s$, $h^B = h^A + 2K_u/M_s$, where, D is a component of the demagnetizing matrix along the field direction. In particular, for a very long and thin samples aligned parallel to the field, the corresponding demagnetizing component D become zero. The macroscopic magnetization curves $m(h, x_A) = x_A m_A(h) + x_B m_B(h)$ in this case are shown in **Fig.2**.

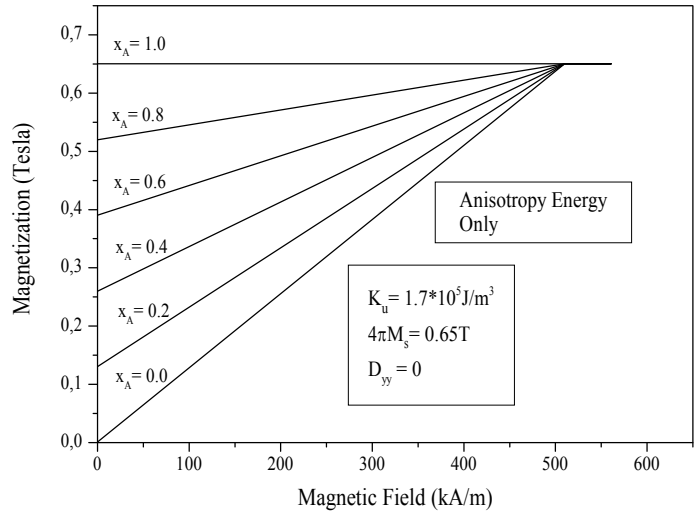

**Fig.2.** Magnetization curves caused by the magnetic anisotropy energy calculated without magnetostatic energy contribution at zero demagnetization factor.

As it is shown in **Fig.2**, the magnetization process occurs by jump at an extremely small field and the magnetization within the variant A immediately takes its saturation value: $m_A(h) = M_s$. Then, it remains constant during the further magnetic field growth. The magnetization of the variant B will occur linearly: $m_B(h) = M_s(h/h^B)$ until its saturation value is achieved at $h = h^B = 2K_u / M_s$.

The second (magnetostatic) contribution to the free energy is strongly dependent on the volume fractions and can be explained on the basis of the general micromagnetic theory according to Eq.16. This term is caused by the magnetostatic energy strongly dependent on the volume fractions. In this case both the magnetic anisotropy and the magnetostatic energy contributions produce a dramatic change in the magnetization curve behavior, shown in **Fig.3**.

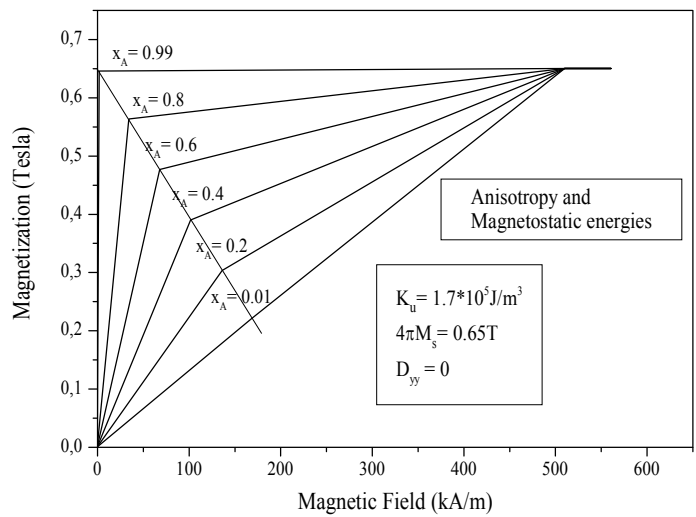

**Рис.3**. Magnetization curves calculations taking into account both the magnetic anisotropy and the magnetostatic energy contributions at zero demagnetizing factor along the magnetic field.

In this case, the magnetization process will occur linearly in both variants in the small field region until the full saturation in the variant A happens at $h = h_A^S(x_A)$.

The saturation field $h = h_A^S(x_A)$ of the variant A is a singular point where, the magnetization curves have a jump-like change of their slope. Its value linearly depends on the volume fraction $x_A$. For the larger field values, the magnetization curves show the linear field dependence until their full saturation of the variant B will occur at $h = h_B^S = 2K_u/M_s$

For our calculations we have used the following material parameters typical for 5M martensitic phase of NiMnGa: the magnetic anisotropy constant $2K_u/M_s = 0.66T$, the saturation magnetization value $4\pi M_s = 0.65T$ and the demagnetizing factor *D = 0*.

In our further publications these results will be used for the different theoretical calculations of the magnetic driving forces, magnetic field induced strain effects and some other interesting properties of FMSMA's.

**Acknowledgments** Present investigation has been supported from the National Budget Support Program No 6541230 "Support of the Favorite Scientific Direction Development" of Ukraine.

**References**

[1] K. Ullakko, J. Mater. Eng. Perform. 5 (1996)405-409.
[2] S.J. Murrey, M. Marioni, S.M. Allen, R.C. O'Handley, Appl. Phys. Lett. 77 (2000) 886-888.
[3] A. Sozinov, A.A. Likhachev, N. Lanska, K. Ullakko, Appl. Phys. Lett. 80 (2002) 1746-1748.
[4] K. Ullakko, J.K. Huang, C. Kantner, R.C. O'Handley, V.V. Kokorin, Appl. Phys. Lett. 69 (1996) 1966-1968.
[5] K. Ullakko, J.K. Huang, V.V. Kokorin, R.C. O'Handley, Scripta Mater. 36 (1997) 1133-1138.
[6] R.C. O'Handley, J. Appl. Phys. 83 (1998) 3263-3270.
[7] R.D. James, R.Tickle, M. Wuttig, Mater. Sci. Eng. A 273-275 (1999) 320-325.
[8] A.A. Likhachev, K. Ullakko, Eur. Phys. J. B 2 (1999) 1-9.
[9] A.A. Likhachev, K. Ullakko, Phys. Lett. A 275 (2000) 142-151.
[10] A.A. Likhachev, K. Ullakko, J. Phys. IV 11 (2001) Pr8-293-Pr8-298.
[11] A.A. Likhachev, A. Sozinov, K. Ullakko, Proc. SPIE 5387 (2004) 128-136.
[12] A.A. Likhachev, A. Sozinov, K. Ullakko, Mater. Sci. Eng. A 378(1-2) (2004) 513-518.
[13] A.A. Likhachev, Materials Science Forum Vols. 738-739 2013 pp 405-410.
[14] A.A. Likhachev, Chem. Met. Alloys, 2014, **6,** p.p.183-187.
[15] A.A. Likhachev, A. Sozinov, K. Ullakko, Mech. Mater. 38(5-6) (2006) 551-563.
[16] A.A. Likhachev, Mater. Sci. Forum 738-739 (2013) 405-410.
[17] O. Heczko, J. Magn. Magn. Mater. 290-291 (2005) 787-794.
[18] L. Straka, O. Heczko, J. Magn. Magn. Mater. 290-291 (2005) 829-831.
[19] U. Gaitzsch, H. Klauß, S. Roth, L. Schultz, J. Magn. Magn. Mater. 324 (2012) 430-433.
[20] Z. Li, Y. Zhang, C. Esling, X. Zhao, L. Zuo, Acta Mater. 59 (2011) 3390-3397.
[21] O. Heczko, J. Kopeček, L. Straka, H. Seiner, Mater. Res. Bull. 48 (2013) 5105-5109.